\documentclass{AIMS}
 \usepackage{amsmath}
\usepackage{lfmath} 
\usepackage{citesort}
\usepackage{amssymb}

\def\currentvolume{X}
 \def\currentissue{X}
  \def\currentyear{200X}
   \def\currentmonth{XX}
    \def\ppages{X--XX}

\theoremstyle{definition}

\title[Hidden downscale energy cascade  in QG turbulence]
      {Is the subdominant part of the energy spectrum due to downscale energy cascade hidden in quasi-geostrophic turbulence?}

\author[Eleftherios Gkioulekas and Ka Kit Tung]{}

\subjclass{Primary: 76FXX; Secondary:60G60}
 \keywords{2d turbulence,  QG turbulence}

\email{lf@amath.washington.edu; tung@amath.washington.edu}
\numberwithin{equation}{section}
\thanks{}

\begin{document}
\maketitle

\centerline{\scshape Eleftherios Gkioulekas}
\medskip
{\footnotesize
 \centerline{Department of Mathematics}
  \centerline{University  of Central Florida}
   \centerline{Box 161364}
   \centerline{Orlando, FL 32816-1364, USA}
} 

\medskip

\centerline{\scshape  Ka Kit Tung }
 \medskip

  {\footnotesize \centerline{Department of Applied Mathematics }
  \centerline{University of Washington } \centerline{Box 352420} \centerline{ Seattle, WA, 98195-2420, USA } }
 \medskip

 \medskip

\begin{abstract}
In systems governing two-dimensional turbulence, surface quasi-geostrophic turbulence, (more generally $\alpha$-turbulence), two-layer quasi-geostrophic turbulence, etc., there often exist two conservative quadratic quantities, one ``energy''-like and one ``enstrophy''-like.  In a finite inertial range there are in general two spectral fluxes, one associated with each conserved quantity. We derive here an inequality comparing the relative magnitudes of the ``energy'' and ``enstrophy'' fluxes for finite or infinitesimal dissipations, and for hyper or hypo viscosities. When this inequality is satisfied, as is the case of 2D turbulence,where the energy flux contribution to the energy spectrum is small,  the subdominant part will be effectively hidden.  In sQG turbulence, it is shown that the opposite is true: the downscale energy flux becomes the dominant contribution to the energy spectrum.  A combination of these two behaviors appears to be the case in 2-layer QG turbulence, depending on the baroclinicity of the system.
\end{abstract}

\section{Introduction}

The characteristic feature of two-dimensional turbulence is that there are two conserved quantities, kinetic energy and enstrophy. This led Kraichnan \cite{article:Kraichnan:1967:1}, Leith \cite{article:Leith:1968}, and Batchelor \cite{article:Batchelor:1969} to conjecture that there will exist two inertial ranges, one located upscale of the spectral region of injection and another on the downscale side of injection.  In the upscale side, it is assumed that there is only an upscale flux of energy, and no flux of enstrophy.  On the downscale side, likewise, there is only a downscale flux of enstrophy, and no flux of energy. One then uses a dimensional analysis argument to calculate the energy spectrum $E (k)$, where it is assumed that in each inertial range $E (k)$ depends only on the corresponding single  flux and the wavenumber $k$. The same type of argument was used in the energy cascade of three-dimensional turbulence \cite{article:Kolmogorov:1941,article:Kolmogorov:1941:1,article:Batchelor:1947}. Although three-dimensional turbulence also has two conserved quantities, energy and helicity, one has the option to inject energy without injecting helicity.  In two-dimensional turbulence it is not possible to inject energy without injecting enstrophy and vice versa, because the two quantities are related.

Initial efforts to simulate the enstrophy cascade yielded confusing reports of various numerical slopes. Consequently, alternative theories have been proposed over the past 30 years to explain them \cite{article:Saffman:1971,article:Polyakov:1993,article:Moffatt:1986}.  Recently, in carefully set up simulations, it was shown that it is possible to obtain the enstrophy cascade in agreement with the KLB theory \cite{article:Alvelius:2000,article:Kaneda:2001,article:Falkovich:2002}. A numerical simulation with very good diagnostics has  shown that the inverse energy cascade can  be obtained accordingly \cite{article:Vergassola:2000}.  There are also however many papers that question the universality of these results \cite{article:Bowman:2003,article:Bowman:2004,article:Gurarie:2001,article:Gurarie:2001:1,article:Danilov:2003}.  A review can be found in  Tabeling \cite{article:Tabeling:2002} and Gkioulekas and Tung \cite{submitted:Gkioulekas:7}.

Further confusion has resulted from efforts to explain the observed energy spectrum of the atmosphere with the KLB theory. Observations show that there is a robust energy spectrum with slope $-3$ which transitions at large wavenumbers into slope $-5/3$ \cite{article:Gage:1979,article:Nastrom:1986,article:Jasperson:1984,article:Gage:1984}. In the KLB theory, on the other hand, one expects that at small wavenumbers the energy spectrum will have slope $-5/3$ from the inverse energy cascade, which will then  transition at the forcing wavenumber, into a $-3$ slope from the direct enstrophy cascade.  The apparent contradiction between these two predictions has led to various explanations and debate \cite{article:Dewan:1979,article:VanZadt:1982,article:Lilly:1989,article:Lindborg:1999,article:Lindborg:2005,article:Lindborg:2006}.

It was conjectured by Tung and Orlando \cite{article:Orlando:2003} that the observed atmospheric energy spectrum results from the downscale cascade of enstrophy and energy injected at the large  scales by baroclinic instability and dissipated at the smallest length scales.  If $\gn_{uv}$ is the downscale enstrophy flux and $\gee_{uv}$ is the downscale energy flux,  it was suggested that there is a downscale energy cascade and a downscale enstrophy cascade that coexist  on the downscale side of injection. Then,  their separate contributions to the energy spectrum would give the latter a compound spectral shape, with a $-3$ slope transitioning to a shallower $-5/3$ slope as the wavenumber increases. The transition from $-3$ slope to $-5/3$ slope occurs at the transition wavenumber $k_t$. The order of magnitude of $k_t$ can be estimated by dimensional analysis and it is given by  $k_t \approx \sqrt{\gn_{uv}/\gee_{uv}}$.  

General Circulation  models have been shown to be capable of reproducing the Nastrom-Gage spectrum in agreement with observations \cite{article:Mahlman:1999,article:Hamilton:2001,article:Skamarock:2004}. Although the nature of the nonlinear interactions which give rise to the downscale energy flux changes from quasi-geostrophic to stratified three-dimensional in the mesoscales, as far as the energy spectrum $E(k)$ is concerned it is the existence of a downscale energy flux which gives rise to the  $k^{-5/3}$ slope, regardless of the character of the motion.   The recent interest, typified by Lindborg \cite{article:Lindborg:2005}, in understanding the $k^{-5/3}$ slope in terms of three-dimensional stratified turbulence is justified, since it is necessary to account for length scales less than 100km in wavelength  where the quasi-geostrophic assumption fails and vortex stretching becomes important.  It is the view of the authors that  it is equally important to understand why the quasi-geostrophic model is capable of supporting a downscale energy cascade with $k^{-5/3}$ scaling, because one also has to account for the transition of the spectral slope from $-3$ to $-5/3$, which occurs between $1000$km and $700$km, and the upper part of the $-5/3$ slope below $700$km.  These are presumably still within the range where quasi-geostrophy is valid.


Tung and Orlando \cite{article:Orlando:2003} have  demonstrated numerically that a two-layer quasi-geostrophic channel model with thermal forcing, Ekman damping, and hyperdiffusion  can also reproduce the $k^{-3} \goto k^{-5/3}$ compound spectrum. The resolution of these simulations goes down to 100km in wavelength.  The diagnostic shown in figure 7 of Tung and Orlando \cite{article:Orlando:2003},  shows both the constant downscale energy and enstrophy fluxes coexisting in the same inertial range.  Recent measurements and data analysis by Cho and Lindborg \cite{article:Lindborg:2001} have confirmed the existence of a downscale energy flux and estimate $\gn_{uv} \approx 2\times 10^{-15} \text{s}^{-3}$ and $\gee_{uv} \approx 6\times 10^{-11}  \text{km}^2  \text{s}^{-3}$. From these estimates we find the mean value of the transition scale $k_t = \sqrt{\gn_{uv}/\gee_{uv}} \approx 0.57\times 10^{-2}  \text{km}^{-1}$ and $\gl_t = 2\pi/k_t \approx 1\times 10^3  \text{km}$ which has the correct order of magnitude.

 This theory of a combined downscale enstrophy cascade and  downscale energy cascade is contrary to the widely accepted misconception that the argument by \Fjortoft \cite{article:Fjortoft:1953} forbids a downscale energy flux in two-dimensional turbulence altogether, and through the isomorphism theorem of Charney \cite{article:Charney:1971} also in quasi-geostrophic turbulence. Various  aspects of this misconception have been clarified by Merilees and  Warn \cite{article:Warn:1975}, Tung and Welch \cite{article:Welch:2001}  and Gkioulekas and Tung \cite{submitted:Gkioulekas:2}. 

As has been pointed out by previous authors \cite{article:Borue:1994,article:Eyink:1996}, as long as the dissipation terms at large-scale and small scales have finite viscosity coefficients, and the inertial ranges exist, the downscale enstrophy flux will be accompanied by a small downscale energy flux, and the upscale energy flux will be accompanied by a small upscale enstrophy flux. Dimensional analysis arguments are premised on the assumption that these additional fluxes can be ignored, consequently the energy spectrum predictions obtained by such arguments are valid only to leading order.  While ignoring subleading effects  can be justified for strictly two-dimensional turbulence, we will argue in this paper that for models of quasi-geostrophic turbulence, such as the two-layer model,  the subleading contributions can be  important in the inertial range and cannot be safely ignored. Predicting the form of these subleading corrections requires a subtle mathematical argument that goes beyond dimensional analysis.

Such an argument was given previously by the authors \cite{article:Tung:2005,article:Tung:2005:1} as follows:  the  subleading fluxes are associated with a subleading downscale energy  cascade and a subleading inverse enstrophy cascade that contribute  \emph{linearly} to the total energy spectrum in addition to the  dominant contributions. 
 To establish this association, we begin with the observation that the underlying MSR theory \cite{article:Rose:1973,article:Phythian:1977} of two-dimensional turbulence is a linear theory and allows two homogeneous solutions that can be linearly superimposed: a solution that represents a downscale enstrophy cascade and a solution that represents a downscale energy cascade.  Then, we argue that the existence of a small downscale energy flux, when the viscosity $\nu$ is finite, implies the presence of a linear combination of \emph{both} solutions in inertial range.  Furthermore, due to the linearity of the MSR theory, the two  homogeneous solutions  are independent of each other, so the downscale energy cascade is independent of the downscale enstrophy flux $\gn_{uv}$ and the downscale enstrophy cascade is independent of the downscale energy flux $\gee_{uv}$. 

It follows that, in the downscale inertial range, the total energy spectrum $E(k)$ has the following three contributions:
\begin{equation}
E(k) = E_{uv}^{(\gee)} (k) + E_{uv}^{(\gn)} (k) + E_{uv}^{(p)} (k), \;\forall k\ell_0 \gg 1,
\label{eq:energy}
\end{equation}
with $E_{uv}^{(\gee)} (k)$, $E_{uv}^{(\gn)} (k)$  given by
\begin{align}
E_{uv}^{(\gee)} (k) &= a_{uv}\gee_{uv}^{2/3} k^{-5/3}\cD_{uv}^{(\gee)} (k\ell_{uv}^{(\gee)}) \\
E_{uv}^{(\gn)} (k) &=  b_{uv}\gn_{uv}^{2/3} k^{-3}[\chi + \ln (k\ell_0)]^{-1/3} \cD_{uv}^{(\gn)} (k\ell_{uv}^{(\gn)}).
\end{align}
Here $E_{uv}^{(\gee)} (k)$, $E_{uv}^{(\gn)} (k)$ are the contributions of the downscale energy and enstrophy cascade, with $\cD_{uv}^{(\gee)}$ and $\cD_{uv}^{(\gn)}$ describing the dissipative corrections. We use the logarithmic correction of Kraichnan \cite{article:Kraichnan:1971:2}, adjusted by the constant $\chi$ of Bowman \cite{article:Bowman:1996} for the contribution of the enstrophy cascade. We have also assumed without explicit justification that we may ignore the possibility of intermittency corrections to the subleading downscale energy cascade. For the downscale enstrophy cascade intermittency corrections have been ruled out by Eyink \cite{article:Eyink:2001}.  For the downscale energy cascade we conjecture that intermittency corrections are small for the same reasons as in three dimensional turbulence. The scales $\ell_{uv}^{(\gee)}$, $\ell_{uv}^{(\gn)}$ are the dissipation length scales for the downscale energy and enstrophy cascade. Finally,  $E_{uv}^{(p)} (k) $ is the contribution  from the effect of forcing and the sweeping interactions. The latter can become significant via the violation of statistical homogeneity caused by the  boundary conditions (see  Ref. \cite{article:Gkioulekas} for details).

 Thus, in the inertial range, if it exists, where  the effect of forcing and dissipation can be ignored, the energy spectrum will take the simple form in the downscale range:
\begin{equation}
E(k) \approx a_{uv} \gee_{uv}^{2/3}k^{-5/3} + b_{uv}\gn_{uv}^{2/3}k^{-3} [\chi + \ln (k\ell_0)]^{-1/3}.
\end{equation}
We see that the energy spectrum will take the slope of $-3$ for small $k$ , and $-5/3$ for large $k$ .  The transition from one slope to the other occurs at $k_t$ , given by $\gee_{uv} k_t^2 \sim \gn_{uv}$.

In two-dimensional turbulence, the fluxes $\gee_{uv}$ and $\gn_{uv}$ are constrained by an inequality that was communicated to us by Sergey Danilov \cite{article:Tung:2005:1}.  This constraint implies that the contribution of the downscale energy cascade to the energy spectrum is overwhelmed by the contribution of the downscale enstrophy cascade and cannot be seen visually on a plot.  This result was conjectured earlier by  Smith \cite{article:Smith:2004} who claimed that the downscale energy cascade can never have enough flux to move the transition wavenumber $k_t$  into the inertial range.  The two-layer model is a different dynamical system than the two-dimensional Navier-Stokes equations, and the validity of the Danilov inequality in the two-layer model is not obvious \cite{article:Tung:2005:1,article:Tung:2004}.

In the present paper we will show that in the two-layer quasi-geostrophic model, when the Ekman dissipation coefficient $\nu_E$ is below a critical value, then the Danilov inequality will be satisfied.  We will also argue that the asymmetric presence of Ekman damping on the bottom layer but not the top layer 
may cause  the violation of the Danilov inequality  for larger values of the Ekman dissipation coefficient. In this case,  the top layer has more enstrophy than the bottom layer, as is realistic in the atmosphere, and provided that the difference in enstrophy between the two layers is large enough, the downscale energy cascade will be made observable in the energy spectrum. We derive specific conditions on how large this difference needs to be in order for the Danilov inequality to be violated for some wavenumbers $k$.  The simulation of Tung and Orlando \cite{article:Orlando:2003} has  shown that it is possible to have an observable downscale energy cascade, which implies a violation of the Danilov inequality.  The role of the argument in this paper is to explain how and why this can happen, given that it  is a surprising and very unexpected result. We also hope that our paper will raise renewed interest in understanding the phenomenology of the two-layer quasi-geostrophic model.

In general, the exposure of the downscale energy cascade requires that the dynamics of the underlying model must be baroclinic.  The role of asymmetric dissipation in inducing baroclinicity was discussed previously by Arbic \cite{thesis:Arbic:2000,submitted:Arbic:1}. We will show that in the surface quasi-geostrophic model, which represents the most extreme case of baroclinicity, the downscale energy cascade becomes completely dominant. An immediate implication of our argument is that the existence of an extensive observable $k^{-5/3}$ in the energy spectrum of the atmosphere has the physical interpretation that the atmosphere is very baroclinic. 

It should be noted that the main results of the present paper rely on the statement that  the transition wavenumber $k_t$ from $-3$ scaling to $-5/3$ scaling is given by $k_t \approx \sqrt{\gn_{uv}/\gee_{uv}}$. This statement follows from the superposition principle, but it doesn't require it and can  also  be established with  the closure model of L'vov and Nazarenko \cite{article:Nazarenko:2006}.  Furthermore, the dimensional estimate $k_t \approx \sqrt{\gn_{uv}/\gee_{uv}}$ for the transition wavenumber $k_t$ has been confirmed by Tung and Orlando \cite{article:Orlando:2003}. Possible criticisms of the superposition principle were discussed in a previous paper \cite{submitted:Gkioulekas:7}, but they are not relevant to the argument of the present paper. Extending the superposition principle to the two-layer QG model (the main focus of this paper), which has three conservation laws instead of just two,  remains an open theoretical  question.

The paper is organized as follows.  The Danilov inequality is reviewed in section 2 where we make some simple generalizations.  Its implications for two-dimensional turbulence  are discussed in section 3. The surface quasi-geostrophic model is discussed in section 4 and  the two-layer model in section 5.  Conclusions and some further remarks are given in section 6.

\section{The Danilov inequality in one-layer models}

 The governing equation  for a wide range of one-layer hydrodynamic models takes the form:
\begin{equation}
\frac{\partial \zeta}{\partial t}  + J (\psi, \zeta ) = {\mathcal D} + {\mathcal
F},
\label{eq:conslaw}
\end{equation}
where ${\mathcal F}$ is the forcing and ${\mathcal D}$ is the dissipation and $\zeta = -\mathcal{L} \psi$. Here, $\mathcal{L}$ is a linear isotropic operator involving the derivatives with respect to the horizontal coordinates. For a general combination of hyper- and hypo-diffusion:
\begin{equation}
{\mathcal D} = -\nu_0 (-\Delta)^p \zeta - \nu_1 (-\Delta)^{-h} \zeta,
\end{equation} 
with 
$p$, $h$, positive integers. $p=1, h=0$ yields the combination of molecular viscosity and Ekman damping.  

For 2D turbulence, $\mathcal{L}$ is given by $\mathcal{L}=-\Delta$, where $\Delta$ is the Laplacian operator and the streamfunction $\psi$ is related to the 2D nondivergent
velocity as
 \begin{equation}
(u, v) = \left( -\pderiv{\gy}{y},  \pderiv{\gy}{x} \right).
\end{equation}  
For barotropic QG turbulence, also known as Charney-Hasegawa-Nima (CHM) turbulence \cite{article:Charney:1948,article:Mima:1978,article:Kodama:1979}, $\mathcal{L}$ is given instead by  $\mathcal{L}=-\Delta +\lambda^2$, where  $\lambda^2$ is a given positive constant. Another interesting family of one-layer models are the $\ga$-turbulence models where $\mathcal{L}=\Lambda^\alpha$ with $\Lambda \equiv (-\Delta)^{1/2}$.  The case $\ga=1$ corresponds to surface quasi-geostrophic turbulence (SQG) which is an extreme baroclinic model, and not a barotropic model like 2D turbulence or CHM turbulence.

\subsection{Conservation laws}

Let $\snrm{f}$ be the spatial and ensemble average of $f(x,y)$ defined as 
\begin{equation}
\snrm{f}\equiv \iint \avg{f(x,y)} \;dxdy.
\end{equation}
There are two inviscid quadratic invariants for \eqref{eq:conslaw}, which are: 
\begin{align}
A&= (1/2)\snrm{ (-\psi \zeta)},\\
B &= (1/2)\snrm{ \zeta^2}.
\end{align}
Note that $B$ is  conserved for all linear operators $\cL$, whereas the conservation law of $A$  requires that $\cL$ be self-adjoint, i.e. 
\begin{equation}
\snrm{f (\cL g)} = \snrm{(\cL f) g}.
\end{equation}
 For example in 2D turbulence it is well-known that, 
\begin{align}
E &\equiv  (1/2) \snrm{(u^2 + v^2)}  = (1/2)\snrm{ \vert \nabla \psi\vert^2 } \\
&= (1/2)\snrm{ (-\psi \zeta)}  \equiv A,
\end{align}
is the kinetic energy of the 2D fluid, and 
\begin{equation}
G \equiv (1/2)\snrm{ \zeta^2} \equiv B,
\end{equation}
 is the enstrophy. In section 4 we see that $A$ and $B$ have a different physical interpretation in SQG turbulence.

 The  spectra $A (k)$ and   $B (k)$ of the conserved quadratics $A$ and $B$ are defined as
\begin{align}
A (k) &= \frac{1}{2} \frac{d}{dk}\snrm{-\gy^{<k}\gz^{<k}}, \\
B (k) &= \frac{1}{2} \frac{d}{dk}\snrm{(\gz^{<k})^2},
\end{align}
with $\gy^{<k}$ and $\gz^{<k}$ the streamfunction and vorticity fields with all the Fourier wavenumbers greater than $k$ in magnitude filtered out. The relationship, $\zeta =-\mathcal{L} \psi$, translates into the spectral relationships in the Fourier space 
\begin{equation}
\hat \zeta ({\mathbf k}) = L(\vert{\mathbf k}\vert) \hat \psi ({\mathbf k}),~~
B (k) = L(k) A (k),
\end{equation}
where 
\begin{equation}
\hat \psi ({\mathbf{k}}) = \iint \psi ({\mathbf{x}})
e^{-i{\mathbf{k}} \cdot {\mathbf{x}}} d\mathbf{x}.
\end{equation}
  We will assume that $L(k)>0$, so that both $A(k)$ and $B(k)$ are positive. Furthermore, we will assume that $L(k)$ is a monotonically increasing function of $k$. In 2D turbulence, $L(k)=k^2$; in CHM turbulence, $L(k)=k^2+\lambda^2$; in $\ga$-turbulence, $L(k)=k^{\alpha}$, and in SQG, $L(k)=k$.

Furthermore, $A (k)$ and $B (k)$ satisfy the following spectral equations:
\begin{align}
\frac{\partial A (k)}{\partial t}  + \frac{\partial \Pi_A (k)}{\partial k}  &=
-D_A (k) + F_A (k)\label{eq:consA}\\
\frac{\partial B (k)}{\partial t}  + \frac{\partial \Pi_B (k)}{\partial k}  &=
-D_B (k) + F_B (k)\label{eq:consB}.
\end{align}
Here, $D_A (k)$ and $D_B (k)$ are the spectral dissipations rate of $A(k)$ and $B(k)$, respectively, with 
\begin{align}
D_B (k) &= L(k) D_A (k), \\
D_A (k) &= [\nu_0 k^{2p} + \nu_1 k^{-2h} ] A (k)> 0,
\end{align}
for a combination of hyper- and hypo-viscosities. Furthermore, $F_A (k)$ and $F_B (k)$ are the spectra of forcing also related by $F_B (k) = L(k) F_A (k)$, and, $\Pi_A (k)$ and $\Pi_B (k)$
are the spectral fluxes of $A$ and $B$.  
The Leith \cite{article:Leith:1968} constraint on the fluxes generalizes to
\begin{equation}
\pderiv{\Pi_B (k)}{k} = L(k) \pderiv{\Pi_A (k)}{k}.
\end{equation}
The conservation laws for $A$ and $B$ arise from the following boundary conditions on $\Pi_A (k)$ and $\Pi_B (k)$: 
\begin{align}
\lim_{k\goto 0^{+}} \Pi_A (k) &= \lim_{k\goto +\infty} \Pi_A (k) = 0 \\
\lim_{k\goto 0^{+}} \Pi_B (k) &= \lim_{k\goto +\infty} \Pi_B (k) = 0.
\end{align}

\subsection{The Danilov inequality}

Assuming that the injection (forcing) of $A$ and $B$ occurs in $[k_1,k_2]$, then at steady state, we have, from \eqref{eq:consA} and \eqref{eq:consB}:
\begin{align}
\Pi_A (k) &=\int_k^{+\infty} D_A (q)\; dq, \text{ for } k > k_2\\
\Pi_B (k) &= \int_k^{+\infty} D_B (q)\; dq, \text{ for } k > k_2\\
\Pi_A (k) &= -\int_0^k D_A (q)\; dq, \text{ for } 0<k<k_1\\
\Pi_B (k) &= -\int_0^k D_B (q)\; dq, \text{ for } 0<k<k_1,
\end{align}
since $F_A (k) =0$ and $F_B (k) = 0$ for $0<k<k_1$ and $k > k_2$. For wavenumbers $k>k_2$, we have  therefore
\begin{align}
L(k)\Pi_A (k) - \Pi_B (k) 
&= \int_k^{+\infty} [L(k) - L(q)] D_A (q) \; dq < 0.
\end{align}
Similarly, for wavenumbers $0<k<k_1$, we have:
\begin{align}
L(k)\Pi_A (k) - \Pi_B (k) &=  - \int_0^k  [L(k) - L(q)] D_A (q) \; dq < 0.
\end{align}
Consequently, for all wavenumbers $k \in (0, k_1) \cup (k_2, +\infty)$ not in the forcing range, we have:
\begin{equation}
L(k)\Pi_A (k) - \Pi_B (k) < 0.
\label{eq:danilov1}
\end{equation}
This inequality was brought to our attention by Danilov (2004, personal communication) for the case of 2D turbulence.  We shall see later in this paper that although this inequality has a trivial mathematical derivation in one-layer models, it becomes  nontrivial for the two-layer QG model.  In all cases, what is interesting is the physical understanding that follows from studying this inequality, and not the mathematical derivations per se.

Previously, \Fjortoft \cite{article:Fjortoft:1953} and  Eyink \cite{article:Eyink:1996} derived a similar, but looser, bound, for the downscale energy flux $\Pi_E$ in two-dimensional turbulence: $\Pi_E (k) < \eta_0 /k^2$, involving the total rate of enstrophy injection $\eta_0$. This looser inequality is often used to show \cite{book:Salmon:1998} that in two-dimensional turbulence with an infinite downscale range, the energy flux $\Pi_E (k)$, vanishes.   For the case of small but finite viscosity where the downscale spectral range is finite, the energy flux does not vanish. 

The significance of the inequality  \eqref{eq:danilov1} is that it decides whether the transition wavenumber $k_t$ is within the inertial range, thus making a transition from the leading cascade to the subleading cascade observable in the energy spectrum $E(k)$. Whether this happens depends on the baroclinicity of the system, as we will show below by considering  three different cases: two-dimensional turbulence  which is completely barotropic, SQG turbulence which is completely baroclinic, and the two-layer QG model which lies in between with respect to baroclinicity.

 \subsection{Integral form of the Danilov inequality}

 The Danilov inequality can be rewritten in an integral form, by noting first that
 \begin{equation}
  \Pi_B (k) =  \int_0^k \pderiv{\Pi_B (q)}{q} \; dq = \int_0^k  L(q) \pderiv{\Pi_A (q)}{q}  \; dq = L(k)\Pi_A (k) - \int_0^k \pderiv{L (q)}{q} \Pi_A (q) \; dq
 \end{equation}
 where the Leith constraint is used.  It follows that
 \begin{equation}
 \int_0^k \pderiv{L (q)}{q} \Pi_A (q) \; dq = L(k)\Pi_A (k) - \Pi_B (k) < 0, \;\forall k\in (0, k_1)\cup (k_2, +\infty).
 \end{equation}
 Note that the validity of the inequality is not affected by the integral going through the forcing range as long as the endpoint wavenumber $k$ is outside the forcing range  $[k_1, k_2]$. The immediate implication of this inequality is that most of the ``energy'' $A$ has to go to small wavenumbers, provided that $\pderivin{L (q)}{q} >0$.  Unlike the flawed argument by  \Fjortoft \cite{article:Fjortoft:1953}, this argument does not rule out the possibility of a small downscale energy cascade.

Likewise, it can be shown that
 \begin{equation}
 \int_k^{+\infty} \frac{1}{[L (q)]^2} \pderiv{L (q)}{q}\Pi_B (q)> 0, \;\forall k\in (0,k_1)\cup (k_2, +\infty).
 \end{equation}
 A detailed discussion of the integral forms of the Danilov inequality was given in a previous paper \cite{submitted:Gkioulekas:2}.

\section{The case of 2D turbulence}

We begin with the  case of 2D turbulence in finite domain with finite viscosity for the infrared and ultraviolet dissipations. This is a generalization of the theory by Kraichnan \cite{article:Kraichnan:1967:1} of infinite domain with infinitesimal dissipation. The Danilov inequality in the inertial range is given by
\begin{equation}
k^2\Pi_A (k) - \Pi_B (k) < 0.
\end{equation}
In the inertial range on the downscale side of injection,  $\Pi_A (k) \approx \epsilon_{uv}$, and $\Pi_B(k)\approx\eta_{uv}$, provided that we ignore a small dissipative contribution. It follows that  $\epsilon_{uv} k^2 <\eta_{uv}$  for all $k$ in this inertial range. The energy spectrum in \eqref{eq:energy}, valid in the inertial range, can be  rewritten to leading order, omitting the logarithmic correction:
\begin{align}
E(k) &\approx C_1 \gee_{uv}^{2/3} k^{-5/3} + C_2 \gn_{uv}^{2/3} k^{-3} \\
&\approx  C_2 \gn_{uv}^{2/3} k^{-3} \left( 1 + \frac{C_1}{C_2}\fracp{\gee_{uv} k^2}{\gn_{uv}}^{2/3} \right)\\
&\approx C_2 \gn_{uv}^{2/3} k^{-3},
\end{align}
where we use $\epsilon_{uv} k^2 <\eta_{uv}$ and the assumption $C_1\leq C_2$. This sequence of steps is valid \emph{asymptotically}
in the limit of large separation between the forcing scale and the dissipation scale, for wavenumbers $k$ in the inertial range. A similar argument can be made for the inertial range upscale of  injection where we use $\Pi_A (k) \approx -\epsilon_{ir}$ and $\Pi_B(k)\approx -\eta_{ir}$.

It follows that on the downscale side of injection the dominant cascade is the enstrophy cascade with $E(k) \sim k^{-3}$, and on the upscale side of injection the dominant cascade is the inverse energy cascade  with $E(k) \sim k^{-5/3}$. By ``dominant'' we mean that even for finite Reynolds numbers the contributions of the subleading downscale energy cascade and the subleading inverse enstrophy cascade are hidden for \emph{all} the wavenumbers $k$ in the inertial range. A violation of the condition $C_1\leq C_2$ can allow, in principle, a transition to $k^{-5/3}$ scaling very near the neighborhood of the dissipation range.  The Danilov inequality by itself cannot completely rule out such an effect, but so far as we know, it  has never been clearly observed in  numerical simulations of the enstrophy cascade.

This argument supports  the conjecture by  Smith \cite{article:Smith:2004} that in 2D turbulence, on the downscale side of injection, we have   no transition to shallower scaling $E (k) \sim k^{-5/3}$.  His other conjecture, that the same result also holds for the two-layer QG model, is not true in general and will be discussed later in this paper. 

It should be noted that in the foregoing arguments it is \emph{assumed} that an inertial range exists either upscale or downscale of injection. Unlike the case of 3D turbulence, where the downscale energy cascade is very robust, it is well known that in 2D turbulence there are circumstances where the leading inverse energy cascade \cite{article:Tung:2005:1,article:Gurarie:2001,article:Gurarie:2001:1,article:Danilov:2003} or the leading downscale enstrophy cascade \cite{article:Shepherd:2002,article:Bowman:2003,article:Bowman:2004} may fail to appear as expected.  Some of these issues are also relevant to the case of $\alpha$-turbulence \cite{article:Tran:2004}. In general, the failure of cascades is to be attributed to the  absense of a sufficiently strong   large-scale dissipation sink.  Since the observational evidence suggests that cascades exist in atmospheric turbulence, we will simply assume that without further discussion.

\section{The case of SQG turbulence}

There has been considerable confusion over the physical interpretation of the surface quasi-geostrophic model.  Although its mathematical formulation is in the form of a one-layer model, it represents a three-dimensional system that corresponds to the baroclinic limit of the three-dimensional quasi-geostrophic model.  Once that is taken into account, the physical interpretation of the spectra $A(k)$ and $B(k)$ and the physical  implications of the Danilov inequality  have  to be revised. The viewpoint  that we would like to put forth in this section is that whereas two-dimensional turbulence is an extreme case where the enstrophy cascade is completely dominant,  SQG turbulence is the other extreme case where it is the downscale energy cascade that is completely dominant. 

\subsection{3D interpretation of SQG turbulence}

As derived by Charney \cite{article:Charney:1971}, 3D QG flow conserves the 3D potential vorticity $\xi$, which is advected horizontally by the streamfunction $\psi$. Here, both $\psi$ and $\xi$ are 3D fields. For constant Coriolis parameter $f$, the governing conservation  law for $\xi$ takes the form:
\begin{equation}
\frac{\partial \xi}{\partial t} + J (\psi, \xi) = 0,
\end{equation}
with $\xi$ given by
\begin{equation}
\xi = \Delta \psi + \frac{f^2}{\rho_0} \frac{\partial}{\partial z} \left
( \frac{\rho_0}{N^2} \frac{\partial\psi}{\partial z} \right) \equiv
\mathcal{P} \psi,
\end{equation}
where $\rho_0 (z)$ is the ambient air density, and $N^2 (z)$ the Brunt-V\``ais\''al\``a frequency. Here we have omitted the forcing and dissipation terms.The streamfunction $\gy$ is also linked with the potential temperature $\Theta$ via the hydrostatic relation
\begin{equation}
\Theta = \frac{g}{N}\frac{T}{T_0} = \frac{f}{N}\pderiv{\psi}{z}.
\end{equation}
The potential temperature is also governed by a 2D-like equation
\begin{equation}
\frac{\partial \Theta}{\partial t} + J (\psi, \Theta) + \frac{T_0 N^2}{g} w = 0,
\end{equation}
where $w$ is the vertical velocity field.

In SQG the potential vorticity $\xi$ is assumed, a priori, to be identically zero for $z > 0$.  The streamfunction $\psi$ is solved from  $\xi =\mathcal{P} \psi =0$.  With $\rho_0$ and $N^2$ taken to be constants, the horizontal Fourier transform of $\psi (x, y, z, t)$ is obtained as
\begin{equation}
\hat\psi ({\mathbf{k}}, z, t) = \hat\psi_0 ({\mathbf{k}}, t) e^{-\vert
{\mathbf{k}} \vert (N/f) z},
\end{equation}
using the boundedness boundary condition as $z \to \infty$.

Most of the dynamics in this model are occurring at the surface $z=0$, where the boundary condition of vanishing vertical velocity $w$ applied to the potential temperature ($\Theta$) equation leads to:
\begin{equation}
\frac{\partial \Theta}{\partial t} + J (\psi, \Theta) = {\mathcal D} +
{\mathcal F},
\label{eq:temp}
\end{equation}
where $\Theta$ now plays the role of the conserved quantity $\zeta$ in \eqref{eq:conslaw}. Here we have also introduced thermal forcing and dissipation:  ${\mathcal D} = \nu \Delta \Theta$ is the  thermal diffusion, and 
\begin{equation}
{\mathcal F} = Q = \alpha_E (\Theta_0 - \Theta),
\end{equation}
 is the thermal heating in the commonly used form of Newtonian cooling (see Ref. \cite{article:Orlando:2003:1}) which includes a ``forcing'' term $\alpha_E \Theta_0$ and the ``Ekman damping'' term $-\alpha_E \Theta$. This equation is to be solved on a 2D surface $z = 0$.  It has the same form as the vorticity equation for 2D turbulence (e.g. \eqref{eq:conslaw}), except that the spectral relationship between the advected quantity $\Theta$ and the advecting field $\psi$ is given instead by 
\begin{align}
\hat\Theta (\bfk, z,  t) &= \frac{f}{N}\pD{z} \left( \hat\psi_0 (\bfk, t) e^{-| \bfk | (N/f) z} \right) \label{sqgcond}\\
&= -| \bfk | [ \hat\psi_0 (\bfk, t) e^{-| \bfk | (N/f) z} ] \\
&= -| \bfk |  \hat\psi_0 (\bfk, z, t),
\end{align}
which reduces to $\hat\Theta ({\mathbf k},t) = -\vert {\mathbf k}\vert \hat\psi (\bfk,t)$ at $z=0$. Thus SQG corresponds to $L(k)= k$. 
 
It has been shown by Charney \cite{article:Charney:1971}, and more generally by  Tung and Orlando \cite{article:Orlando:2003:1},  that the 3D QG energy density
\begin{equation}
\mathcal{E} \equiv \frac{1}{2} \rho_0\left[\vert \nabla \psi \vert^2 +
\frac{f^2}{N^2} \left ( \frac{\partial \psi}{\partial z} \right)^2
\right],
\end{equation}
is an invariant (i.e. independent of time), \emph{ when integrated over the 3D domain.} $\cE$ is the sum of the kinetic energy density $\cE_K$ and the potential energy density $\cE_P$ which are given by
\begin{align}
\mathcal{E}_K&=(1/2) \rho_0
    (u^2 + v^2) = (1/2) \rho_0 | \nabla \psi|^2 \\
\cE_P &= \frac{1}{2}\rho_0\fracp{f}{N}^2 \left( \pderiv{\psi}{z}\right)^2 = \frac{1}{2} \rho_0 \Theta^2.
\end{align}
For SQG, using Parseval's identity, the energies integrated over the horizontal surface are given by
\begin{align}
E_P &= \snrm{\cE_P} = \frac{1}{2}\gr_0 \snrm{\Theta^2},\\
E_K &= \snrm{\cE_K} = \frac{1}{2}\gr_0 \snrm{| \nabla\psi |^2}  \\
&= \frac{1}{2}\gr_0 \int \avg{(i\bfk \hat\psi (\bfk, t))\cdot (-i\bfk \hat\psi^{\ast} (\bfk, t))} \; d\bfk \\
&=  \frac{1}{2}\gr_0 \int \avg{ |-k \hat\psi (\bfk, t) |^2 } \; d\bfk =   \frac{1}{2}\gr_0 \snrm{\Theta^2} = E_P.
\end{align}
It is thus seen that the kinetic energy density and the available potential energy density, when integrated horizontally, are equipartioned,  and that 
\begin{equation}
2B \equiv \snrm{ \Theta^2}=(E_P+E_K)/\rho_0=E/ \rho_0,
\end{equation}
is the total energy at the lower surface.  The 3D energy is, instead
\begin{align*}
E_{3D} &\equiv \int_0^\infty\snrm{\mathcal{E}}dz = \int_0^\infty \rho_0
dz \snrm{ \Theta^2} \\
&= \int_0^\infty
\rho_0 dz \iint dk_x dk_y \avg{\left \vert \hat\Theta \vert_{z=0} \right 
\vert^2} e^{-2\vert {\mathbf k}\vert(N/f) z}\\
&= \left.\frac{1}{2} \rho_0 \iint dk_x dk_y \frac{f}{N\vert {\mathbf k}\vert} \avg{\hat \Theta
\hat\Theta^\ast\right\vert_{z=0}}\\
&= \frac{1}{2} \rho_0 \iint \frac{f}{N} \avg{-\hat\psi \hat\Theta^\ast
\vert_{z=0}} dk_x dk_y \\
&= \frac{1}{2} \rho_0\frac{f}{N} \left. \snrm{ (-\psi
\Theta\right\vert_{z=0})} = \rho_0(f/N) A,
\end{align*}
 with $A$ defined earlier as $A \equiv (1/2)\snrm{(-\psi \Theta )}.$ Previous authors have made use of the similarity between the form of vorticity equation  \eqref{eq:conslaw} in 2D turbulence and the temperature equation \eqref{eq:temp} in SQG turbulence to identify, by analogy, $A$ as the ``energy'' and $B$ as the ``enstrophy'' \cite{article:Swanson:1994,article:Swanson:1995}.  As pointed out  in  Tung and Orlando \cite{article:Orlando:2003:1} and also here, $2B$ is the total energy integrated over the lower surface, and includes kinetic plus available potential energy.  The physical interpretation for $A$ was not given, but can now be seen to be the total energy integrated over the 3D domain. There is no potential enstrophy $(\xi^2/2)$ per se in SQG turbulence, because potential vorticity $\xi$ has been taken to be zero identically. Consequently the  flux of potential enstrophy in SQG is exactly equal to zero, thus $k^2 \Pi_{E_{2D}} (k) - \Pi_G (k) > 0$ and $k^2 \Pi_{E_{3D}} (k) - \Pi_G (k) > 0$.

\subsection{The energy spectrum in SQG turbulence and $\alpha$-turbulence}

The argument of the previous section can be extended to the case of SQG turbulence and to the even wider case of $\alpha$-turbulence, which includes both the case of SQG for $\alpha=1$, and the case of 2D for $\ga=2$. Here $L(k)=k^{\alpha}$ and we assume $\ga > 0$. Using the linear superposition principle  \cite{article:Tung:2005} and assuming the existence of inertial ranges, the spectrum of $A(k)$ and $B(k)$ are, in the downscale inertial range:
\begin{align}
A (k) &\approx C_1 (\gee_A)^{2/3} k^{-\frac{7}{3} + \frac{1}{3} \alpha} + C_2
(\gee_B )^{2/3} k^{-\frac{7}{3} - \frac{1}{3} \alpha}\\
B (k) &= \vert k\vert^\alpha A (k).
\end{align}
Here, $\gee_A$ and $\gee_B$ are the constant fluxes on the downscale side of the forcing range. The inequality \eqref{eq:danilov1}  becomes $k^\alpha \gee_A <\gee_B$  for all $k$ in the inertial ranges. Consequently, for wavenumbers $k$ in the inertial range, the spectrum $A(k)$ is given by 
\begin{align}
A(k) &\approx C_2 (\gee_B)^{2/3} k^{-\frac{7}{3} - \frac{1}{3}\ga} \left( 1 + \frac{C_1}{C_2}\fracp{\gee_A k^{\ga}}{\gee_B}^{2/3} \right) \\
&\approx C_2 (\gee_B)^{2/3} k^{-\frac{7}{3} - \frac{1}{3}\ga},
\end{align}
again in the limit of large separation between the forcing scale and the dissipation scale and provided that $C_1 \leq C_2$.
Thus, in the downscale range, there is no observable transition and therefore:
\begin{align}
A (k) &\approx C_2 (\gee_B)^{2/3} k^{-\frac{7}{3} - \frac{1}{3} \alpha},\\ 
B(k) &\approx C_2 (\gee_B)^{2/3} k^{-\frac{7}{3} + \frac{2}{3} \alpha}.
\end{align}

For the SQG model, in the downscale inertial range, the energy spectrum is the same as that for $2B(k)$ in $\alpha$-turbulence with $\alpha=1$:
\begin{equation}
E (k) = 2B (k) \approx C_1 \gee_{3D}^{2/3} k^{-1} + C_2 \gee_{2D}^{2/3} k^{-5/3}.
\end{equation}
Here $\gee_{3D}=\gee_A$ is the downscale energy flux of 3D energy $A$ and $\gee_{2D}=\gee_B$ is the downscale energy flux of 2D energy $B$.
From the Danilov inequality we learn that the visible energy spectrum in the inertial range downscale from the injection scale is given by
\begin{equation}
E (k) \approx C_2 \gee_{2D}^{2/3} k^{-5/3}.
\end{equation}
This $k^{-5/3}$ energy spectrum is now predicted by our theory. The flux $\gee_{2D}$  is not the ``enstrophy'' flux but is the 2D flux of 2D energy $2B = E$ at the $z=0$ layer.  Numerical simulations have reproduced this energy cascade and appear to indicate a small deviation from the $k^{-5/3}$ slope due to conjectured intermittency corrections \cite{article:Vergassola:2004}.


\subsection{The relevance of SQG for the troposphere and the Nastrom-Gage spectrum} 

In the troposphere, because of fast vertical mixing, the potential vorticity is almost homogenized and is very close to zero.  It takes a sharp jump to a large positive potential vorticity in the stratosphere across the tropopause.  So if we take the tropopause to be the ``surface'' and take $z$ as positive pointing downward, we have a situation approximated by SQG for those wavenumbers which are large compared to the distance between the tropopause and the ground.  For these wavenumbers, which satisy $k(N/f)z\gg 1$ in Eq \eqref{sqgcond}  for all $z$ between $0$ and the tropopause height of $H$, the ground appears to be infinitely far.  Since the SQG solution applies in this case, and since we have predicted that the energy spectrum has the slope of $-5/3$ due to downscale flux of 2D potential energy, this then may explain the observed Nastrom-Gage spectrum for the wavenumbers larger than $1/(HN/f)\sim 1/1000$km, since $H\sim 10$km and $N/f$ is typically $100$ in the troposphere.  For smaller $k$, the presence of the lower ground needs to be taken into account, and a nonlinear Eady problem needs to be solved.  Two layer models, on the other hand, are a good approximation for the scales which ``see'' both the tropopause and the ground, and therefore are  good candidates for simulating the $-3$ part of the Nastrom-Gage spectrum.

 It is important to note that when viewed this way, the presence of the $-5/3$ slope in the Nastrom-Gage spectrum is suggesting that the troposphere is very baroclinic in the vertical.  This situation cannot be approximated by a two-dimensional turbulence model.  When using a two-layer model to approximate this baroclinic situation, the two layers must be very different. In the next section we argue that this can be achieved with asymmetric Ekman damping. A difference between SQG and a two-layer model is that the downscale flux in the former is finite even as viscosity goes to zero from above.  In the latter the finite downscale energy flux needs to be maintained by a finite viscosity, because of the well-known absence of anomalous dissipation in two-layer models. 

\section{The case of two-layer QG turbulence}

The results in the previous sections  demonstrate that QG turbulence can exhibit a variety of behaviors.  Barotropic models  usually  possess an energy spectrum with $-3$ spectral slope, as in 2D turbulence, while SQG turbulence, which is baroclinic (with its exponential decay with height), is expected to have a spectrum with a $-5/3$ slope, in the downscale inertial range.  Two-layer QG models have both a barotropic and a baroclinic component \cite{article:Salmon:1978,article:Salmon:1980,book:Salmon:1998}, and we therefore expect a mixture of $-3$ and $-5/3$ slopes depending on the degree of baroclinicity. In terms of the Danilov inequality, we expect that it will hold for small wavenumbers and fail for large wavenumbers for sufficiently baroclinic systems.   

A relatively realistic  two-layer model applicable to studying atmospheric turbulence in the troposphere was adopted in discussed by Tung and Orlando \cite{article:Orlando:2003}.  In this model forcing is due to thermal heating, which injects energy directly into the baroclinic part of the total energy.  The two-layer fluid sits atop of an Ekman boundary layer near the ground, which introduces Ekman pumping in the lower layer \cite{book:Holton:1979} but {\it not} in the upper layer.  If one artificially adds an identical  Ekman damping in the upper layer  it can be shown that Danilov's inequality applies, and we leave the proof to the interested reader. Consequently, asymmetric Ekman damping is the only possible mechanism that can lead to a violation of the Danilov inequality for quasi-geostrophic models.  

This significance of asymmetric Ekman damping on the dynamical properties of quasi-geostrophic models was also discovered independently by Arbic \cite{thesis:Arbic:2000,submitted:Arbic:1}, who investigated a set of similar but distinct inequalities.  For the more general case of three-dimensional models, there are  additional mechanisms, which act synergistically with asymmetric dissipation.  

For the multiple-layer QG model, the governing equations can be rewritten in  the form of the conservation law with $\zeta=-\mathcal{L}\psi$, if  we make $\mathcal{L}$ into a matrix and $\psi$ into a column vector.  We will discuss the more general  theory of multi-layers in a future paper.  Here, we only wish to explain  why and how the Danilov inequality can fail in the two-layer QG model.

\subsection{The formal setup}

Two-layer QG models conserve potential vorticity in each layer in the absence of forcing and damping.  In the forced-dissipative case, the governing equations read:
\begin{align}
\text{Top layer:}\qquad &\frac{\partial \zeta_1}{\partial t}  + J
(\psi_1, \zeta_1) = {\mathcal D}_1 + {\mathcal F}_1\\
\text{Bottom layer:}\qquad &\frac{\partial \zeta_2}{\partial t} + J (\psi_2, \zeta_2) =
{\mathcal D}_2 + {\mathcal F}_2,
\end{align}
where $$\zeta_1 = \Delta \psi_1 - \frac{k_R^2}{2} (\psi_1 - \psi_2),~
\zeta_2 = \Delta \psi_2 + \frac{k_R^2}{2} (\psi_1 - \psi_2),$$
are the potential vorticity in each layer.  $k_R \equiv (2\sqrt{2} f)/(hN)$ is the Rossby radius of
deformation wavenumber and is taken as a given constant ($h$ is the height).  The dissipation
terms, ${\mathcal D}_i$, include momentum dissipation of relative
vorticity, $\Delta \psi_i$, in each layer, and Ekman damping from the lower
boundary layer:
\begin{equation}
{\mathcal D}_1 = \nu (-\Delta)^{p+1} \psi_1,~~
{\mathcal D}_2 = \nu (-\Delta)^{p+1} \psi_2 - \nu_E\Delta \psi_2.
\end{equation}
The forcing terms can be shown to satisfy 
\begin{equation}
\cF_1 = +\frac{k_R^2}{2f} Q, ~~\cF_2 = -\frac{k_R^2}{2f} Q,
\end{equation}
where $Q$ is the radiative heating term in the temperature equation.

The two inviscid quadratic invariants are the  energy $E$ and the total layer potential enstrophies $G_1$ and $G_2$ given by
\begin{align}
E &\equiv \snrm{\gy_1 \gz_1 + \gy_2 \gz_2} \\
G_1 &\equiv \snrm{\gz_1^2}, \quad G_2 \equiv\snrm{\gz_2^2}.
\end{align}
The energy and enstrophy spectra are defined as
\begin{align}
E(k) &\equiv \spec{\gy_1}{\gz_1}{k} + \spec{\gy_2}{\gz_2}{k},\\
G_1 (k) &\equiv \spec{\gz_1}{\gz_1}{k},\\
G_2 (k) &\equiv \spec{\gz_2}{\gz_2}{k},
\end{align}
and the total enstrophy spectrum $G (k)$ is $G(k) = G_1 (k) + G_2 (k)$. We define the streamfunction spectra 
\begin{align}
U_1 (k) &\equiv \spec{\gy_1}{\gy_1}{k},\\
U_2 (k) &\equiv \spec{\gy_2}{\gy_2}{k},\\
C (k) &\equiv \spec{\gy_1}{\gy_2}{k},
\end{align}
and also $U(k) \equiv U_1 (k) + U_2 (k)$. It is easy to show the triangle inequality $2C(k) \leq U(k)$.   The energy and enstrophy spectra are related with the streamfunction spectra via
\begin{align}
E (k) &= (k^2 +k_R^2/2) U(k) - k_R^2 C(k)\\
G(k) &= (k^4+k^2 k_R^2 + k_R^4/2) U(k) - k_R^2 (2k^2 - k_R^2) C(k).
\end{align}
From the standpoint of examining the Danilov inequality, it is most convenient to work with the streamfunction spectra.  However, following Salmon \cite{article:Salmon:1978,article:Salmon:1980,book:Salmon:1998}, for physical understanding it is useful to  work with the energy and enstrophy spectra. Furthermore, it is helpful to  distinguish between barotropic energy and baroclinic energy as follows: Let $\gy \equiv (\gy_1 + \gy_2)/2$ and $\gt \equiv (\gy_1 - \gy_2)/2$. So, $\gy_1 = \gy+\gt$ and $\gy_2 = \gy-\gt$. Now we define three spectra $E_K (k)$, $E_P (k)$, and $E_C (k)$ in terms of $\gy$ and $\gt$:
\begin{align}
E_K (k) &\equiv 2k^2 \spec{\gy}{\gy}{k}, \\
E_P (k) &\equiv 2(k^2 + k_R^2) \spec{\gt}{\gt}{k}, \\
E_C (k) &\equiv 2k^2 \spec{\gy}{\gt}{k}.
\end{align}
Here $E_K (k)$ is the barotropic energy spectrum and $E_P (k)$ the baroclinic energy spectrum. It is easy to show that the definitions are self-consistent, i.e. $E (k)=E_K (k)+E_P (k)$. The relation between the energy spectra and the enstrophy spectra can be now written in terms of two constraint equations:
\begin{align}
G_1 (k) &= (1/2)[k^2 E_K (k) + (k^2+k_R^2) E_P (k)] +  (k^2+k_R^2) E_C (k) \\
G_2  (k) &= (1/2)[k^2 E_K (k) + (k^2+k_R^2) E_P (k)] -  (k^2+k_R^2) E_C (k).
\end{align}
Thus we see that the physical interpretation of $ E_C (k)$ is that it represents the difference in potential enstrophy distribution between the two layers, and it is given by
\begin{equation}
E_C (k) = \frac{G_1 (k) - G_2 (k)}{2(k^2+k_R^2)}.
\end{equation}

\subsection{Controlled necessary  condition}

The right-hand-side of the Danilov inequality reads
\begin{align}
k^2 \Pi_E (k) - \Pi_G (k) &= \int_k^{+\infty} dq\; (k^2 D_E (q) - D_G (q))\equiv \int_k^{+\infty} dq\; \gD (k,q),
\end{align}
so the key question is whether $ \gD (k,q)$ is positive or negative
for wavenumbers $k<q<k_{max}$. Here $k_{\max}$ is either the truncation wavenumber in the numerical model, or, in the theoretical case of infinite resolutions, is the hyperviscosity dissipation wavenumber, beyond which the spectral
enstrophy dissipation rate becomes negligible.

With some straightforward but tedious calculations, it can be shown that the dissipation rate $D_E (k)$ for the total energy, and  the dissipation rate $D_G (k)$ for the total potential enstrophy are given by
\begin{align}
D_E (k) &= 2\nu k^{2p+2} U(k) + 2\nu_E k^2 U_2 (k) \\
D_G (k) &=  2\nu k^{2p+2} E(k) + \nu_E k^2 [(2k^2 + k_R^2) U_2 (k)-  k_R^2 C(k)],
\end{align}
thus it follows that
\begin{align}
\gD (k,q) &= k^2 D_E (q) - D_G (q) \\
& = 2\nu q^{2p+2} [k^2 U(q) - E(q)] + \nu_E k_R^2 q^2 C(q) \\ 
& \quad+ \nu_E q^2 (2k^2 - 2q^2 -  k_R^2) U_2 (q) \\
&=   2\nu q^{2p+2} [(k^2-q^2) U(q) + (k_R^2/2)(2C(q) - U(q))] \\
& \quad + \nu_E k_R^2 q^2 (C(q)-U_2 (q)) + 2\nu_E q^2 (k^2-q^2) U_2 (q).
\end{align}
The first and third terms in this expression are always negative.  Consequently, a \emph{necessary}
condition for \emph{violating} the Danilov inequality is that the second term has to be positive, i.e. $C(q)-U_2 (q)\geq 0$.  Otherwise, if $C(q)-U_2 (q)\leq 0$, then the Danilov inequality will be satisfied. A physical interpretation of this condition will be given in section \ref{sec:layerinterp}.

A controlled sufficient condition to satisfy the Danilov inequality can be derived in terms of the physical parameters of the problem by noting that $C(q)-U_2 (q) \leq (1/2)(U_1 (q) - U_2(q))$. It follows that
\begin{align}
\gD (k,q) &= 2\nu q^{2p+2} [(k^2-q^2) U(q) + (k_R^2/2)(2C(q) - U(q))]  \\ 
& \quad + \nu_E k_R^2 q^2 (C(q)-U_2 (q)) + 2\nu_E q^2 (k^2-q^2) U_2 (q)\\
&\leq 2\nu q^{2p+2} [(k^2-q^2) U(q) + (k_R^2/2)(2C(q) - U(q))]  \\ 
& \quad + \nu_E k_R^2 q^2 (1/2) (U_1 (q)-U_2 (q)) + 2\nu_E q^2 (k^2-q^2) U_2 (q) \\
&= [2\nu q^{2p+2} (k^2-q^2) + (1/2) \nu_E k_R^2 q^2] U_1 (q) \\ 
& \quad +  [2(\nu q^{2p+2}+\nu_E q^2)(k^2-q^2) - (1/2) \nu_E k_R^2 q^2] U_2 (q) \\
&\leq [2\nu q^{2p+2} (k^2-q^2) + (1/2) \nu_E k_R^2 q^2] U_1 (q).
\end{align}

Here, we have used the inequality $2C(k) \leq U(k)$ to eliminate the $(k_R^2/2)(2C(q) - U(q))$ term. We have also eliminated the $U_2 (q)$ term because it is unconditionally negative. This leads to the following controlled \emph{sufficient} condition to \emph{satisfy} Danilov's inequality:
\begin{equation}
\nu_E < 4\nu k^{2p}_{\max} \fracp{k_{\max}}{ k_R}^2.
\end{equation}
 Equivalently, a \emph{necessary} condition to \emph{violate} Danilov's inequality is 
 \begin{equation}
 \nu_E > 4\nu k^{2p}_{\max}\fracp{k_{\max}}{k_R}^2.
 \end{equation}

 It is interesting to note that  in the numerical simulation of the two-layer model the algorithm adopted by  Tung and Orlando \cite{article:Orlando:2003} for determining the magnitude of the hyperviscosity
coefficient is $\nu_E \gg \nu k^{2p}_{\max}$, for all but the last
twenty wavenumbers $k$ in the dissipation range.  Tung and Orlando \cite{article:Orlando:2003} obtained an energy spectrum with the compound slope configuration and the  transition wavenumber $k_t$ occured in the inertial range downscale from injection in agreement with the condition $k_t \approx \sqrt{\eta_{uv}/\epsilon_{uv}}$, thus implying a violation of Danilov's inequality.

\subsection{An uncontrolled necessary and sufficient condition}

The question now arises: is it possible to derive a sufficient condition to violate the Danilov inequality of the form $\nu_E k_R^2 \geq \gL \nu k_{\max}^{2p+2}$ for some universal constant $\gL$? So far as we know, this is not possible.  However, it is possible to derive an \emph{uncontrolled} necessary and sufficient condition for violating the Danilov inequality.

We begin with defining 
\begin{align}
2C (q) &= \gl (q) U(q)\\
U_2 (q)&= u(q) U(q)
\end{align}
 Here $0\leq u (q) \leq 1$ and $-1\leq \gl (q) \leq 1$. We may thus rewrite everything in terms of $U(q)$ by employing  
\begin{align}
C(q)-U_2 (q) &= (\gl (q)/2 -u(q))U(q)\\
2C(q) - U(q) &= (\gl (q)-1) U(q)
\end{align}
Then we can rewrite $\gD (k,q)$, as follows:
\begin{align}
\gD (k,q) &= 2\nu q^{2p+2} [(k^2-q^2) U(q) + (k_R^2/2)(2C(q) - U(q))]  \\ 
& \quad + \nu_E k_R^2 q^2 (C(q)-U_2 (q)) + 2\nu_E q^2 (k^2-q^2) U_2 (q)\\
&= q^2 U(q) [-\nu q^{2p}(2(q^2 - k^2) + k_R^2 (1-\gl (q))  \\ 
& \quad +  \nu_E k_R^2 (\gl (q)/2 - u(q)+2(k/k_R)^2 u(q)-2(q/k_R)^2 u(q))].
\end{align}
It easy to see that a sufficient condition to get $\gD (k,q) \geq 0$ is
\begin{equation}
\nu_E k_R^2 (\gl (q)/2 - u(q)+2(k/k_R)^2 u(q)-2(q/k_R)^2 u(q)) \geq \nu q^{2p}(2(q^2 -k^2) + k_R^2 (1-\gl (q))).
\end{equation}
The necessary condition to violate the Danilov inequality $C(q)-U_2 (q) \geq 0$, which was derived previously, implies that $\gl (q)/2 -u(q) \geq 0$. Provided that we assume the stronger condition 
\begin{equation}
\gl (q)/2 - u(q)+2(k/k_R)^2 u(q)-2(q/k_R)^2 u(q)\geq 0,
\label{eq:necesscond}
\end{equation}
 we may rewrite our sufficient condition as:
\begin{equation}
\frac{\nu_E k_R^2}{\nu q^{2p}} \geq \frac{2(q^2 -k^2)+ k_R^2 (1-\gl (q))}{\gl (q)/2 - u(q)+2(k/k_R)^2 u(q)-2(q/k_R)^2 u(q)}.
 \label{eq:suffcond}
\end{equation}
To violate the Danilov inequality at wavenumber $k$, this sufficient condition must hold for all $q$ such that $k<q<k_{max}$.
Since the numerator is always positive, the condition \eqref{eq:necesscond} is in fact a stronger  necessary condition for violating the Danilov inequality.

\subsection{Physical interpretation of necessary conditions}
\label{sec:layerinterp}

We would like now to discuss the plausibility of the necessary condition $\gl (q)/2 -u(q) > 0$ and the stronger  necessary condition \eqref{eq:necesscond}.
 To this end, we rewrite these conditions equivalently in terms of the physical energy  and enstrophy spectra.

It is easy to write $U_1 (k)$, $U_2 (k)$, and $C(k)$ in terms of $E_K (k)$, $E_P (k)$, and $E_C (k)$:
\begin{align}
U_1 (k) &= \specp{\gy+\gt}{\gy+\gt}{k}\\
&= \spec{\gy}{\gy}{k} + 2\spec{\gy}{\gt}{k} + \spec{\gt}{\gt}{k}\\
&= \frac{E_K (k)}{2k^2} + \frac{E_P (k)}{2(k^2+k_R^2)} + \frac{E_C (k)}{k^2},
\end{align}
and with a similar argument we find
\begin{align}
U_2 (k) &= \frac{E_K (k)}{2k^2} + \frac{E_P (k)}{2(k^2+k_R^2)} - \frac{E_C (k)}{k^2}\\
C (k) &=  \frac{E_K (k)}{2k^2} - \frac{E_P (k)}{2(k^2+k_R^2)}.
\end{align}
We may thus write $\gl (k)$ and $u(k)$ in terms of $E_K (k)$, $E_P (k)$, and $E_C (k)$:
\begin{align}
\gl  (k)&= \frac{2C (k)}{U (k)} = 2\;\frac{\displaystyle \frac{E_K (k)}{2k^2} - \frac{E_P (k)}{2(k^2+k_R^2)}}{\displaystyle \frac{E_K (k)}{k^2} + \frac{E_P (k)}{(k^2+k_R^2)}} \\
&= \frac{(k^2+k_R^2) E_K  (k)- k^2 E_P (k)}{(k^2+k_R^2) E_K (k) + k^2 E_P (k)},\\
\end{align}
and
\begin{align}
u(k) &= \frac{U_2 (k)}{U (k)} = \frac{\displaystyle \frac{E_K (k)}{2k^2} + \frac{E_P (k)}{2(k^2+k_R^2)} - \frac{E_C (k)}{k^2}}{\displaystyle \frac{E_K (k)}{k^2} + \frac{E_P (k)}{(k^2+k_R^2)}}\\
& = \frac{1}{2}\frac{(k^2+k_R^2) E_K (k) + k^2 E_P (k) - 2 (k^2+k_R^2) E_C (k)}{(k^2+k_R^2) E_K (k) + k^2 E_P (k)}.
\end{align}
and the necessary condition $\gl (q)/2 -u(q) \geq 0$ can now be rewritten as
\begin{equation}
\gl (q)/2 -u(q) = \frac{(q^2+k_R^2) E_C (q) - q^2 E_P (q)}{(q^2+k_R^2) E_K (q) + q^2 E_P (q)} > 0.
\end{equation}
The denominator is obviously positive, consequently the condition is equivalent to
\begin{equation}
G_1 (q) - G_2 (q) = 2 (q^2+k_R^2) E_C (q) > 2q^2 E_P (q),
\end{equation}
for all $q$ such as $k<q<k_{max}$.

This is a very interesting result.  The requirement, in part, is that there should be more enstrophy on the top layer than the bottom layer, i.e. $G_1 (q)>G_2 (q)$.  It is reasonable to expect this if there is more dissipation on the bottom layer than the top layer.  This is the case for the model we are considering where there is Ekman damping at the bottom layer but not at the top layer.  However, the actual condition also requires that the difference should be larger than the potential energy $E_P (q)$ multiplied with $2q^2$ .  In our model, as has been pointed out by Salmon \cite{article:Salmon:1978,article:Salmon:1980,book:Salmon:1998}, the energy is injected into the system at large scales as baroclinic energy, and most of this energy is converted into barotropic energy near the Rossby wavenumber $k_R$. It is therefore reasonable to expect that this necessary condition will be satisfied for wavenumbers $q>k_R$. 

It should be noted that this is only a necessary condition.  A sufficient condition would require furthermore that inequality \eqref{eq:suffcond} be valid. The  necessary precondition for for the inequality to have the desired direction  is the stronger requirement
\begin{equation}
\frac{\gl(q)/2 - u(q)}{u(q)} > 2\fracp{q}{k_R}^2 - 2\fracp{k}{k_R}^2.
\end{equation}
This condition can be rewritten equivalently as
\begin{equation}
\frac{G_1 (q) - G_2(q) - 2q^2 E_P (q)}{2G_2 (q) + k_R^2 (E_K (q) - E_P(q))} > 2\fracp{q}{k_R}^2 - 2\fracp{k}{k_R}^2.
\end{equation}
and it can be simplified further to give
\begin{multline}
G_1 (q) - \left( 1+ 4\;\frac{q^2-k^2}{k_R^2}\right) G_2 (q) > 2q^2 E_K (q) - 2k^2 (E_K (q) - E_P(q)).
\end{multline}
The constraint $u(q) \geq 0$ implies that
\begin{equation}
2 G_2 (q) + k_R^2 (E_K (q)-E_P(q)) \geq 0,
\end{equation}
and subtracting this inequality from the necessary condition above gives the following simplification:
\begin{equation}
G_1 (q)- (1+4(q/k_R)^2) G_2 (q) > 2q^2 E_K (q),
\end{equation}
for all $q$ such that $k<q<k_{max}$.

This condition, which is also a necessary condition for violating the Danilov inequality, places an even stronger constraint on the difference between the enstrophy between the two layers, which is dependent on the ratio $4(q/k_R)^2$. In the numerical simulations of  Tung and Orlando \cite{article:Orlando:2003}, the variability of  $4(q/k_R)^2$  is relatively small because they only go down to 100km in wavelength. It should be noted, of course, that the quasi-geostrophic model is not valid at length scales much smaller than 100km in wavelength. When three-dimensional effects become relevant, the conservation of enstrophy is violated, and additional means for continuing the downscale energy cascade to smaller scales come into play.

\section{Conclusions and Discussion}

 The classical KLB theory of 2D turbulence relies for its mathematical  simplicity and elegance on two unrealistic assumptions: that the domain is  infinite, and that the Reynolds number approaches infinity. When these two assumptions are relaxed, the situation becomes more complicated.  The downscale enstrophy cascade is accompanied with a hidden downscale energy cascade, and similarly the inverse energy cascade is accompanied with a hidden inverse enstrophy cascade.  This is true as long as the leading cascades themselves exist, which requires the presence of a sufficiently strong dissipation sink at small wavenumbers.  The fluxes associated with the subleading cascades are constrained by the Danilov inequality, and as a result the subleading cascades cannot contribute large enough terms to the energy spectrum to create an observable effect. This picture represents a generalization of the KLB theory to finite inertial ranges and finite dissipations. This situation changes, however, in baroclinic models of quasi-geostrophic turbulence.

The surface quasi-geostrophic model represents an extreme baroclinic case where the entire behavior in the three-dimensional domain is constrained by the behavior of the system at the  $z=0$ layer.  In this model there is no enstrophy, and the dominant feature is the downscale energy cascade.

We have shown that in the two-layer quasi-geostrophic model, the violation of the Danilov inequality is possible only as a result of asymmetric Ekman damping operating on only one of the two-layers.  This creates an imbalance between the amount of enstrophy  accumulated in one layer versus the amount accumulated in the other layer, and the downscale energy cascade will become observable on the condition that this imbalance is sufficiently large. We have derived in the present paper a sufficient condition for \emph{not} violating the Danilov inequality which explains why the  $k^{-5/3}$ spectrum has not been observed in some of the previous simulations of the two-layer model. We have also derived a necessary and sufficient condition for violating the Danilov inequality, but it is an uncontrolled condition.  The numerical simulation by Tung and Orlando \cite{article:Orlando:2003} has confirmed that a double cascade with the transition wavenumber located in the inertial range  can be realized. This can only occur when the Danilov inequality is  violated for some wavenumbers $k$ in the inertial range. The parameterization of the Ekman damping in that simulation does in fact satisfy the necessary condition derived in this paper.

As long as we operate within the framework of multiple-layer models with a finite number of layers, one cannot rule out the alternative theory that the atmospheric energy spectrum might reflect a double downscale cascade of helicity and energy instead of enstrophy and energy (see discussion in section 6.5 of Ref. \cite{book:Moiseev:1999}, and figure 3 of \cite{article:Tsinober:1993}).  However, most of the current debate has been focused on the somewhat mysterious nature of the very extensive and robust  $k^{-5/3}$ spectrum. 

 Our work in the present paper explains why it can be reproduced in numerical simulations that use baroclinic models, while the same effect cannot be realized in simulations of two-dimensional turbulence. On the other hand our work here does not rule out the possibility that the shallower part of the spectrum observed by Nastrom and Gage \cite{article:Gage:1984} over the mesoscales can be due to dynamics other than QG, whether it is barotropic or baroclinic, especially for scales of $100$ km or less (see e.g. Ref. \cite{article:Lindborg:2005,article:Lindborg:2006} with Bousinesq dynamics).  Our present work serves to point out that over the larger scales ($\gtrapprox 600\text{km})$, where the transition to a shallower spectrum occurs, baroclinic QG theory by itself is a viable mechanism for explaining the transition from $-3$ to $-5/3$ slopes. 

Furthermore, as proposed first by Tung and Orlando \cite{article:Orlando:2003}, the downscale energy flux, which is important in explaining the $k^{-5/3}$ energy spectrum over the mesoscales in most theories, originates at larger scales (the synoptic scales). Its contribution to the energy spectrum is hidden for smaller wavenumbers under the $k^{-3}$ part of the spectrum, and then emerges for larger $k$ past the transition scale.  It remains an open question, one that is beyond the scope of this paper,  to understand in detail how this downscale energy flux can be continued into  length scales too small for QG theory to describe, and how it is eventually dissipated.

\section*{Acknowledgements} It is a pleasure to thank Dr. Sergey Danilov for his e-mail correspondance which led us to the development of the ideas in this paper. Constructive comments from an anonymous reviewer are also gratefully acknowledged. The research is supported by the National Science Foundation, under grants DMS-03-27658.

\bibliography{references,references-submit}
\bibliographystyle{lfplain}

\medskip


\medskip

\end{document}